\documentclass[12pt]{article}
\title{ Multicritical Behaviours in One-Dimensional Traffic Flow} 
  
\author{ Abdelaziz Mhirech$^a$$^,$$^*$, Hamid Ez-Zahraouy$^a$ and Assia Alaoui Ismaili$^b$ .}  
\begin{document}
\maketitle
\begin{center}
{\it \small
Universit\'e Mohammed V, Facult\'e des Sciences, B.P. 1014, Rabat, Morocco.\\$^a$D\'epartement de Physique, Laboratoire de Magn\'etisme et de la Physique des Hautes \'energies. .\\ $^b$D\'epartement de Math\'ematiques.}
\end{center}

\abstract{ The effect of the on-ramp  and off-ramp positions $i_1$ and $i_2$, respectively, on the one dimensional-cellular automaton  traffic flow behaviour, is investigated numerically. The on-ramp and off-ramp  rates at $i_1$ and $i_2$ are $\alpha_0$ and $\beta_0$, respectively. However, in the open boundary conditions, with injecting  and extracting rates $\alpha$ and $\beta$ and using parallel dynamics, several phases occur, depending on the position of $i_1$ by respect to $i_2$. Namely, low density phase (LDP), intermediate density phase (IDP), plateau current phase (PCP) and high density phase (HDP). Furthermore, critical, tricritical and multicritical behaviours take place in the  $(i_1,\alpha_0)$ phase diagrams.}\\\\
Pacs number :  05.40.-a, 05.50. +q, 64.60.Cn, 82.20.wt
\------------------------------------------\\
$^*$corresponding author e-mail address: mhirech@fsr.ac.ma

\newpage
\section{Introduction}
\setlength{\parskip}{.2in} 
The field of transport have attracted increased attention in recent years [1-4]. This interest is due primarily to the fact that transportation problems are related to the global behaviour of systems with many elements interacting at short distances, such as the vehicles travelling on the streets or informations which travel over the internet network.\\ In particular, the investigation of open traffic systems with on- and  off-ramps is quite popular  at the moment [5-10]. One reason for this is the impact of the understanding of varying the different flow rates in order to optimise the total flow or trip times.\\ Among the different methods of investigation and simulation of highway traffic flow, asymmetric simple exclusion process (ASEP) is the most promising [11-13]. Indeed, ASEP is the simplest driven diffusive system where particles on a one-dimensional lattice hop with asymmetric rates under excluded volume constraints.\\The effect of on- and off-ramps have been largely studied by several researchers in the traffic flow field [5,11]. Our principal aim in this paper is to bring an explanation on the following point: let us suppose that we have a highway running in an urban conglomeration and that there are an access from urban conglomeration to the highway and an exit from the highway. We want to understand where the access and exit positions must be localised in order to maximize the flux of cars in the road. In other words,  we want to study the effect of the on-ramp  and off-ramp positions on the one dimensional-cellular automaton  traffic flow behaviour, in the open boundaries case. To this end, we have established the phase diagrams in the ($i_1,\alpha_0$)-plane. Depending on the injecting and extracting rate values, an adequate localization of the on- and off-ramp positions leads to the appearance of new phases, topologies and several multicritical behaviours. Moreover, to compare our results to those where only one off-ramp was taken into account [5], quantitative differences can be understood from the behaviour of average density and phase diagrams for different parameters.\\

The paper is organised as follows: Model and method are given in section 2; section 3 is reserved to results and discussion; the conclusion is presented in section 4.

\section{Model and method}We consider a one-dimensional lattice of length L. Each lattice site is either empty or occupied by one particle. Hence the state of the system is defined by a set of occupation numbers  $\tau_{1}$,$\tau_{2}$,...,$\tau_{L}$, while $\tau_{i} = 1$ ($\tau_{i} = 0$) means that the site i is occupied (empty). We suppose that the main road is single lane, an on-ramp and an off-ramp connect the main road only on single lattice $i_{1}$ for entry and on single lattice $i_{2}$ for way out. During each time interval $\Delta t$, each particle jump to the empty adjacent site on its right and does not move otherwise ($i\neq i_{2}$). $\Delta t$ is an interesting parameter that enables the possibility to interpolate between the cases of fully parallel ($\Delta t =1$) and random sequential ($\Delta t \rightarrow 0$) updates [12]. Particles are injected, by a rate $\alpha \Delta t$, in the first site being to the left side of the road if this site is empty, and particles enter in the road by site $i_{1}$, with a probability $\alpha_{0} \Delta t$ without constraint, if this site is empty. While, the particle being in the last site on the right can leave the road with a rate $\beta \Delta t$ and particles removed on the way out with a rate $\beta_{o} \Delta t$. At site $i_{1}$ ($i_{2}$) the occupation (absorption) priority is given to  the particle which enter in the road (particle leaving the road). Hence the cars, which are added to  the road, avoid any collision.\\
If the system has the configuration $\tau_{1}(t)$, $\tau_{2}(t)$,...,$\tau_{L}(t)$ at time t it will change at time $ t + \Delta t $ to the following:\\ 
For $i=i_{1}$ , 
\begin{equation} \tau_{i}(t+ {\Delta t}/{2}) = 1 \end{equation}  with probability  \begin{equation} q_{i}=\tau_{i}(t)+[\alpha_{0}(1-\tau_{i}(t))-\tau_{i}(t)(1-\tau_{i+1}(t))]\Delta t \end{equation}     
and 
 \begin{equation}   \tau_{i}(t+ \Delta t/2) = 0   \end{equation} 
 with probability $1-q_{i}$.  Where $i_{1}$ and $\alpha_{0}$ denote the position of the entry site and the injection rate, respectively. 
\\   For $i=i_{2}$ ,  
\begin{equation}   \tau_{i}(t+ \Delta t/2) = 1  \end{equation}  
with probability  
\begin{equation} q_{i}=\tau_{i}(t)+[\tau_{i-1}(1-\tau_{i}) -\beta_{0}\tau_{i}(t)]\Delta t \end{equation} 
and  
\begin{equation}   \tau_{i}(t+ \Delta t/2) = 0  
 \end{equation} with probability $1-q_{i}$.  Where $i_{2}$ and $\beta_{0}$ denote the position of the absorbing site and the absorbing rate, respectively.\\ 
 For 1$<$i$<$L with $i\neq i_{1}$ and $i\neq i_{2}$,  \begin{equation}    \tau_{i}(t+ \Delta t) = 1   \end{equation}  
 with probability  
 \begin{eqnarray} q_{i}=\tau_{i}(t)+[\tau_{i-1}(t)(1-\tau_{i}(t))-\tau_{i}(t)(1-\tau_{i+1}(t))]\Delta t  \end{eqnarray}  
 and   
\begin{equation}   \tau_{i}(t+ \Delta t) = 0   \end{equation}   
with probability $1-q_{i}$. \\\\  
 For $i=1$, 
 \begin{equation}   \tau_{1}(t+ \Delta t) = 1   \end{equation} 
  with probability
 \begin{eqnarray}   q_{1}=\tau_{1}(t)+[\alpha(1-\tau_{1}(t))-\tau_{1}(t)(1-\tau_{2}(t))]\Delta t   \end{eqnarray} 
  and 
 \begin{equation}   \tau_{1}(t+ \Delta t) = 0   \end{equation} 
  with probability $1-q_{1}$. \\ 
   For $i=L$,
   \begin{equation}   \tau_{L}(t+ \Delta t) = 1   \end{equation}  
 with probability 
  \begin{eqnarray}   q_{L}=\tau_{L}(t)+[\tau_{L-1}(t)(1-\tau_{L}(t))-\beta\tau_{L}(t)]\Delta t,   \end{eqnarray} 
  and 
  \begin{equation}   \tau_{L}(t+ \Delta t) = 0   \end{equation} 
  with probability $1-q_{L}$.\\
 \section{Results and Discussion}
In our numerical calculations, the rule described above is updated in parallel, $\Delta t=1$, i.e. during one update step the new particle position do not influence the rest and only the previous positions have to be taken into account [13]. During each of the time steps, each particle moves one site unless the adjacent site on its right is occupied by another particle. \\
 In the following, the temporal average of any parameter is computed for  $5\times 10^{4}$ to $10^{5}$ time steps. Starting the simulations from random configurations, the system reaches a stationary state after a sufficiently large number of time steps. In all our simulations, we averaged over $60-100$ configurations. For the update step, we consider two sub steps as has been used  in a previous work [14].\\ As we have mentioned previously, our aim in this paper is to study the effect of the on- and  off-ramps positions $i_1$ and $i_2$, respectively, for different values of $\alpha$, $\beta_{0}$ and $\beta$, on the average density and flux in the chain. The study is made in the open boundary conditions case. The length of the road studied in this paper is L=1000. Moreover, our goal here is to study the cases where $i_1$ is upstream or downstream $i_2$, we then fixed $i_2=500$ and varied $i_1$.\\\\   
The figure 1 gives  the variation of the average density $\rho$ as a function of the injecting rate $\alpha_0$ for several values of the on-ramp position $i_1$ and for  $\alpha = 0.1$, $\beta = 0.1$ and $\beta_0 = 0.4$. However, when the on-ramp is located upstream the off-ramp, the system studied exhibits four phases, namely:  i) The low density phase (LDP), where the average density  increases when increasing the rate of injected particles $\alpha_{0}$.   ii) The intermediate density phase ($IDP_1$) characterised by a smoothly increase of the density.  iii) The plateau  current phase ($PCP_1$) for which  the density and current are constant in a   special interval of $\alpha_0$.  iv) The high density phase ($HDP_1$) in which, for high values of $\alpha_0$, the current decreases and  the density reaches a large value and remains constant. On the other hand, when $i_1$ is located downstream $i_2$, the high density phase disappears. Note that we add index 1 to each phase when this one is specific to the case $i_1<i_2$ and index 2 in the contrary case. In addition, when increasing $i_1$ for a given value of $\alpha_0$, the average density remains constant in the LDP, while it decreases in the $IDP_1$ and PCP then increases in the $HDP_1$. Moreover, an inversion point is located at a special value of $\alpha_0=\alpha_{0i}$ which corresponds to the $PCP_1-HDP_1$ transition. This inversion can be explained as follows:\\
- For $i_1<i_2$: if $\alpha_0<\alpha_{0i}$ and $i_1$ increases, the section between $i_1$ and $i_2$ decreases reducing the average density ($\alpha$ is small). While, if $\alpha_0>\alpha_{0i}$, there is a fast accumulation of particles in the section between $i_1$ and $L$ ($\beta$ being weak) which increases the average density.\\ 
-For $i_1>i_2$: if $\alpha_0$ is small, the increase in $i_1$ reduces the zone $i_1-L$ thus decreasing the average density. Whereas, when $\alpha_0$ is large, the accumulation of particles between $i_1$ and $L$ is compensated by the reduction in the section $i_1-L$. The average density remains thus constant.\\
  In addition, if $i_1 < i_2$, we note that the IDP, which doesn't appear in the model where only the off-ramp is taken into account [5], occurs for the intermediate values of  $\alpha_0$ ($\alpha_{0c1}<\alpha_0<\alpha_{0c2}$). $\alpha_{0c1}$ and $\alpha_{0c2}$ correspond to the transition between $LDP-IDP_1$ and $IDP_1-PCP_1$, respectively. While, the $PCP_1$ arises between two critical values $\alpha_{0c2}$ and $\alpha_{0c3}$ of injecting rate $\alpha_{0}$, where  $\alpha_{0c3}$ corresponds to the $PCP_1-HDP_1$ transition, which coincide with $\alpha_{0i}$. We note that this transition disappears when $i_1$ is located after $i_2$ (Figure 1).\\
In order to have a suitable criterion for determination of the nature of the transition, we identify the first order transition (abrupt transition) by a jump in the average density or by the existence of a peak in the derivative of $\rho (\alpha_{0})$ with respect to $\alpha_{0}$ [13]. This means that the transitions shown in figure 1 are of first order type.\\ Collecting the results illustrated in figure 1, the different regions are given on the phase diagram ($i_1$,$\alpha_{0}$) shown in Figure 2. Beside this, such phase diagram  exhibits the critical and multicritial behaviours. The critical end-points, around which there is no distinction between the phases, are denoted by $CEP$. 
 In order to describe the different entities in the phase diagrams, the Griffiths notations [15] will be adopted: the multicritical point $A^mB^n$ denotes the intersection of m lines of first order and n lines of second order. The $A^3$ indicates then the triple point, which is the intersection of three lines of first order. The critical point, which is the intersection of a line of first order and a line of second order, is denoted by C.  However, the figure 2 shows two multicritical points $A^2B$ and five single critical end-points CEP. Indeed, for a given value of $\alpha_0$, the transitions $IDP_1-PCP_2$ and $PCP_1-IDP_2$ are of the second order, while the transition $HDP_1-IDP_2$ is of the first order. We identify the second order transition by respect to the average density, which is continuous but presents an angular point [12]. 
For low values of $\alpha$, $\beta$ and $\beta_0$ ($\alpha=0.1$, $\beta=0.1$ and $\beta_0=0.1$), the ($i_1,\alpha_0$) phase diagram is presented in figure 3, where all the transitions are of the first order, except that located between $PCP_1-IDP_2$. Moreover, such figure exhibits three critical end-points and one multicritical point of type $A^2B$. The comparison of figures 2 and 3 shows up the effect of $\beta_0$ on the ($i_1,\alpha_0$) phase diagram. Indeed, for intermediate value of $\beta_0$, the $IDP_1$ and $PCP_2$ take place.\\ Furthermore, for a sufficiently large value of $\beta_0$ ($\beta_0=0.8$, as shown in figure 4), the $IDP_1$ disappears, two critical points (C), two triple points ($A^3$) and two end-points (CEP) occur.\\ Finally, to highlight the effect of $\beta$, we present on figure 5 the ($i_1,\alpha_0$) phase diagram for $\alpha=0.1$, $\beta_0=0.1$ and $\beta=0.3$. In this case, the system exhibits one multicritical point $A^2B$, one critical point C, one triple point $A^3$ and two critical end-points CEP. From figures 2, 3 and 5, we deduce that the $IDP_1$ arises for intermediate values of extracting rates $\beta$ or $\beta_0$, in agreement with the previous work [14].\\
 \section{Conclusion}  Using numerical simulations, we have studied the effect of the  on- and off-ramp positions on the traffic flow behaviour of a one dimensional-cellular automaton, with parallel update. When $i_1$ is upstream $i_2$, an inversion point takes place in $\rho(\alpha_0)$, at $PCP_1-HDP_1$ transition. Depending on the values of $\alpha$, $\beta$ and $\beta_0$, the ($i_1,\alpha_0$) phase diagram exhibits different topologies. The transitions between different phases are of the first or second order. Furthermore, the system exhibits a multicritical, critical and critical end-points.\\
  
 \newpage \textit{Figure captions:}
Fig. 1: Average density $\rho$ versus the injection rate $\alpha_{0}$ for $\alpha=0.1$, $\beta=0.1$, $\beta_0=0.4$. The number accompanying each curve denotes the value of $i_1$.
Fig. 2: Phase diagram  ($i_1,\alpha_{0}$) for $\alpha=0.1$, $\beta=0.1$, $\beta_0=0.4$. The solid and dot lines indicate the first and second order transitions, respectively.\\
Fig. 3: Phase diagram ($i_1,\alpha_{0}$) for $\alpha=0.1$, $\beta_{0}=0.1$ and $\beta=0.1$.\\
Fig. 4: Phase diagram ($i_1,\alpha_{0}$) for $\alpha=0.1$, $\beta_{0}=0.8$ and $\beta=0.1$.\\
Fig. 5: Phase diagram ($i_1,\alpha_{0}$) for $\alpha=0.1$, $\beta_{0}=0.1$ and $\beta=0.3$.
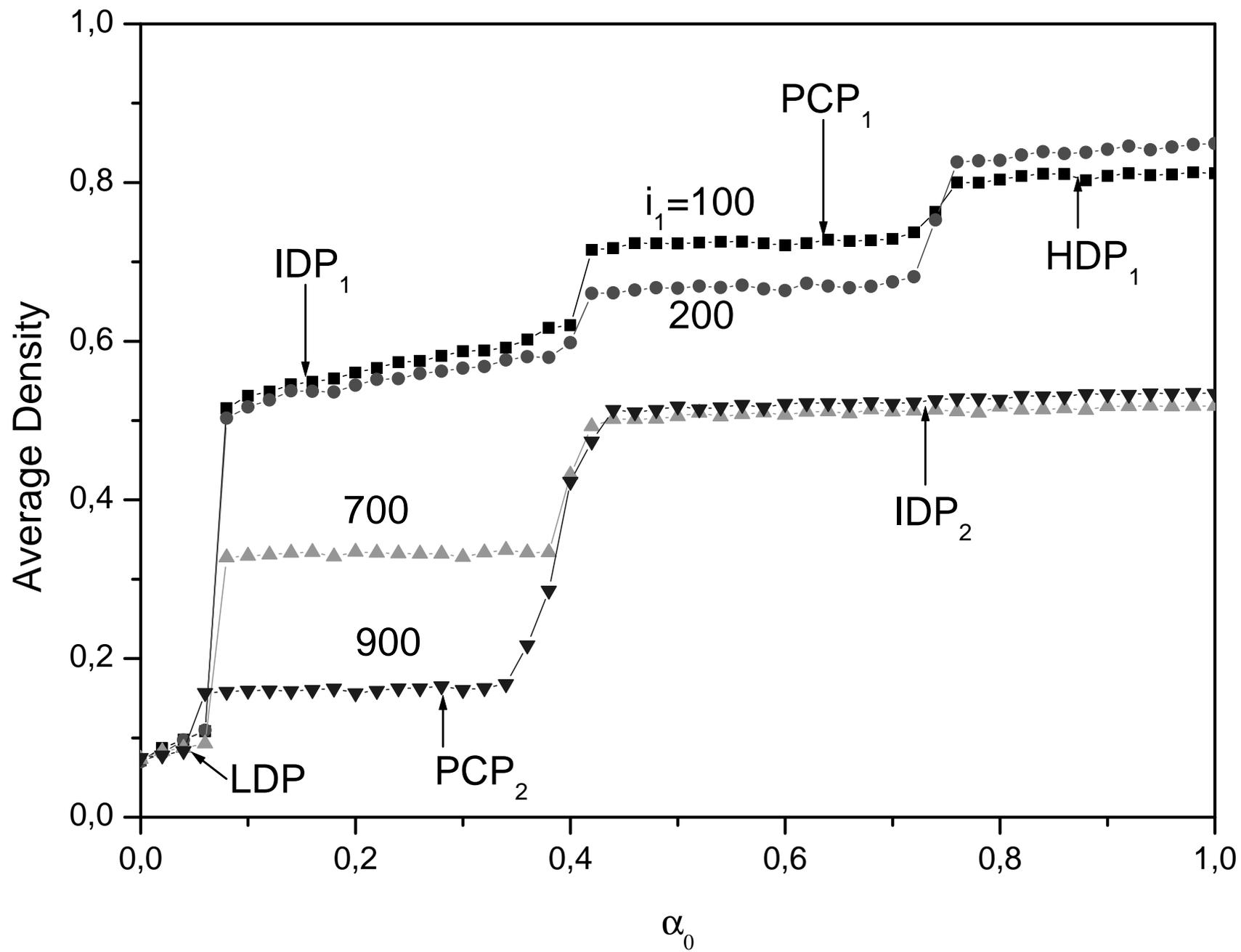
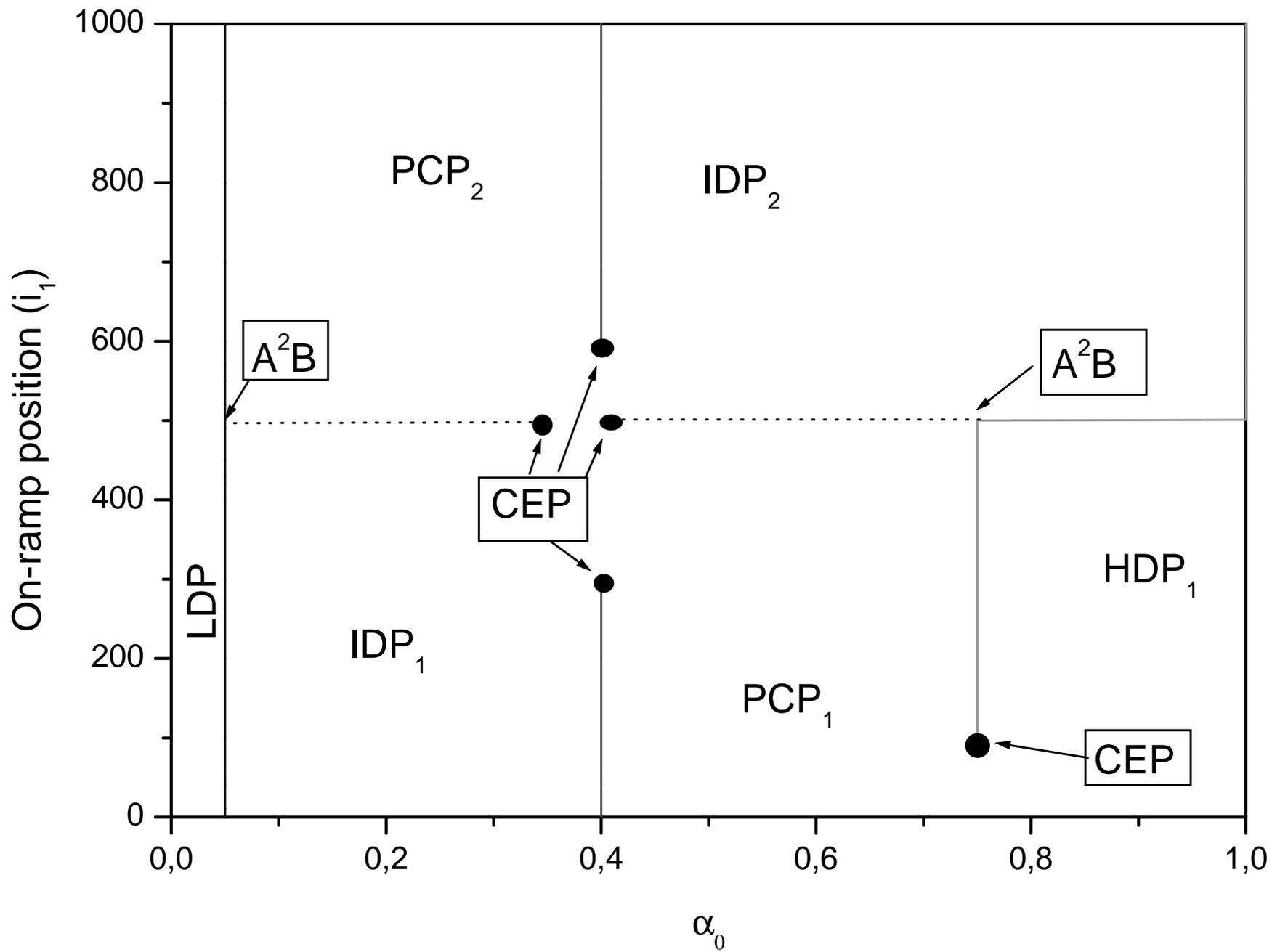
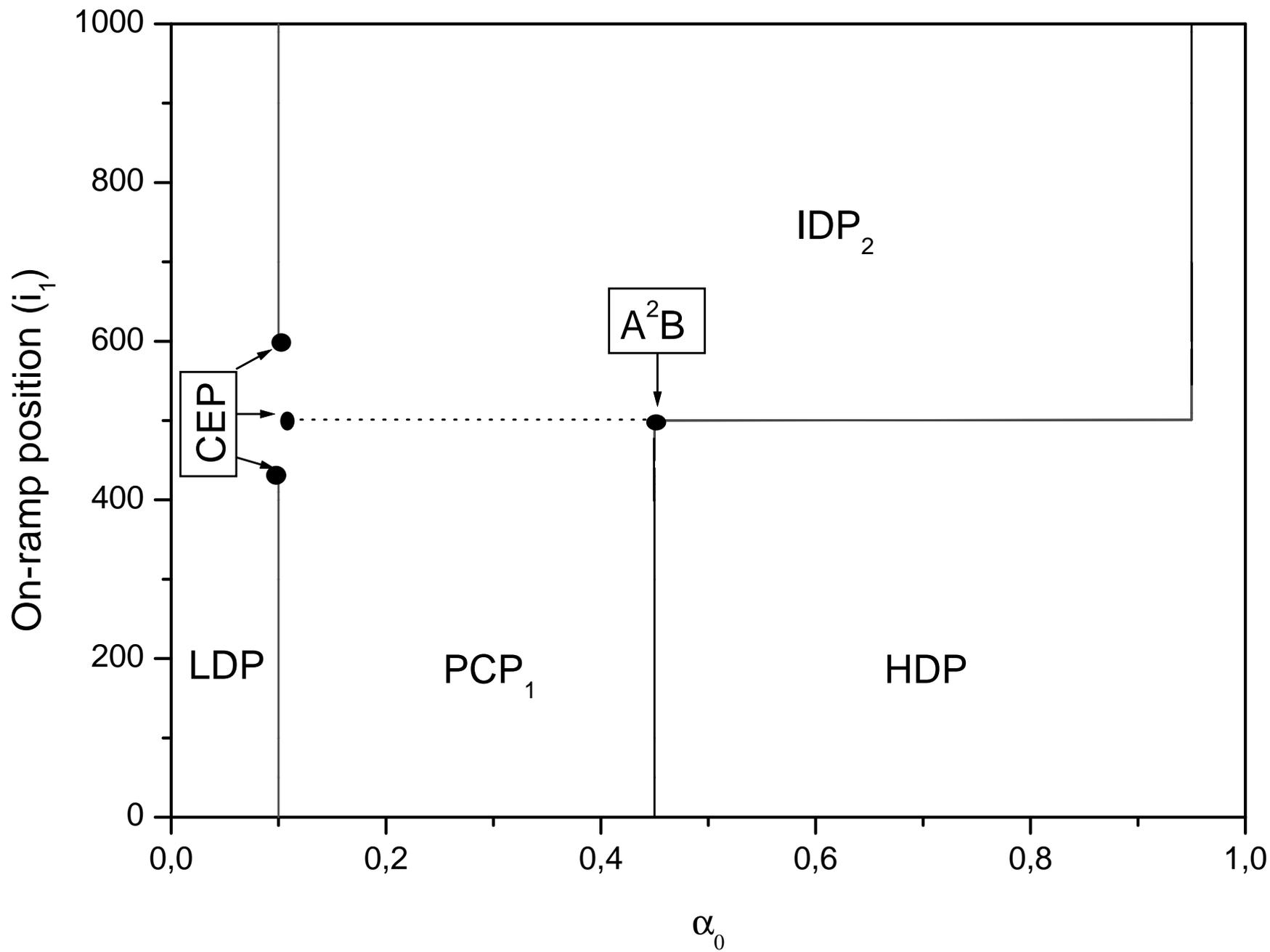
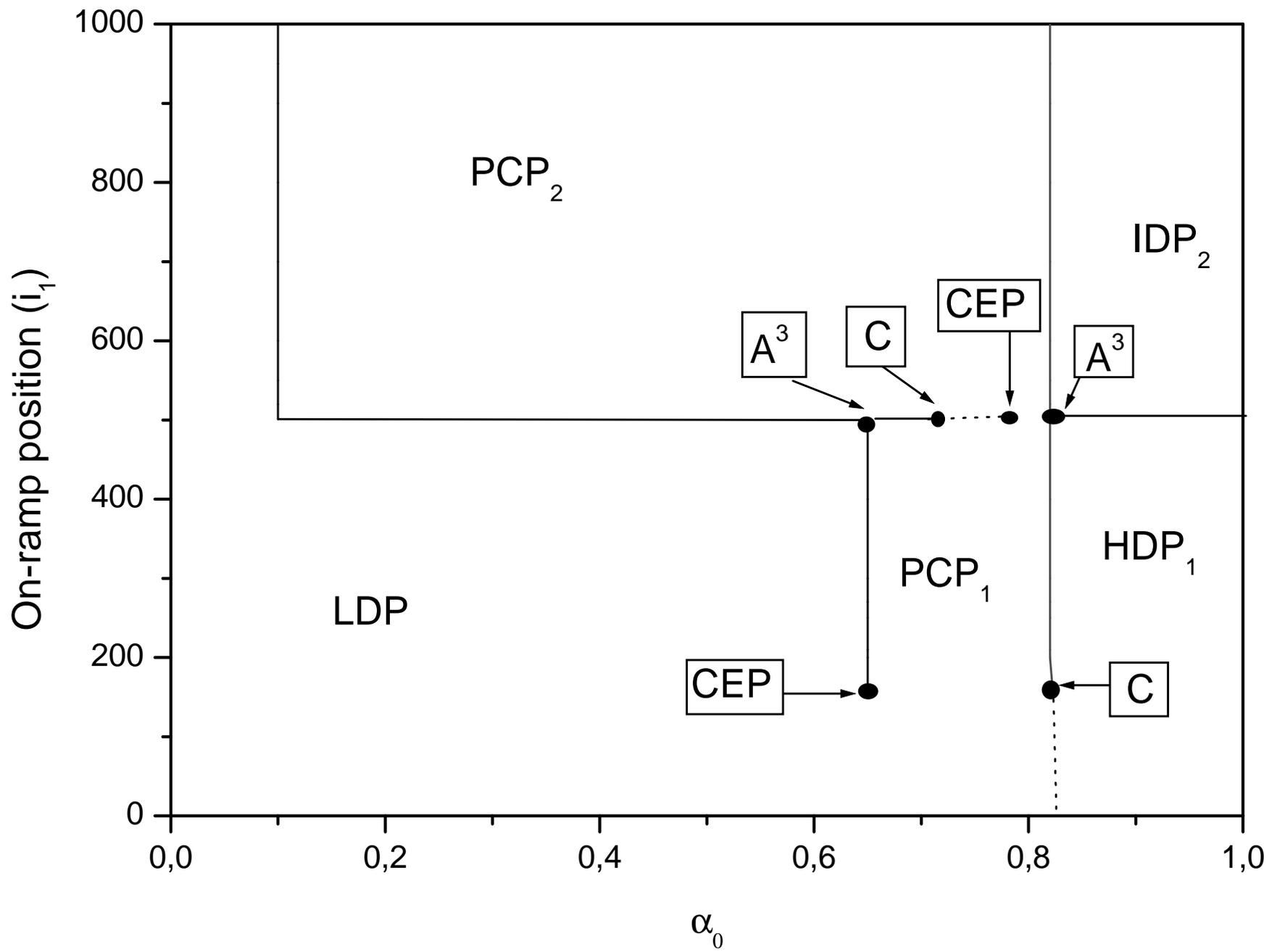
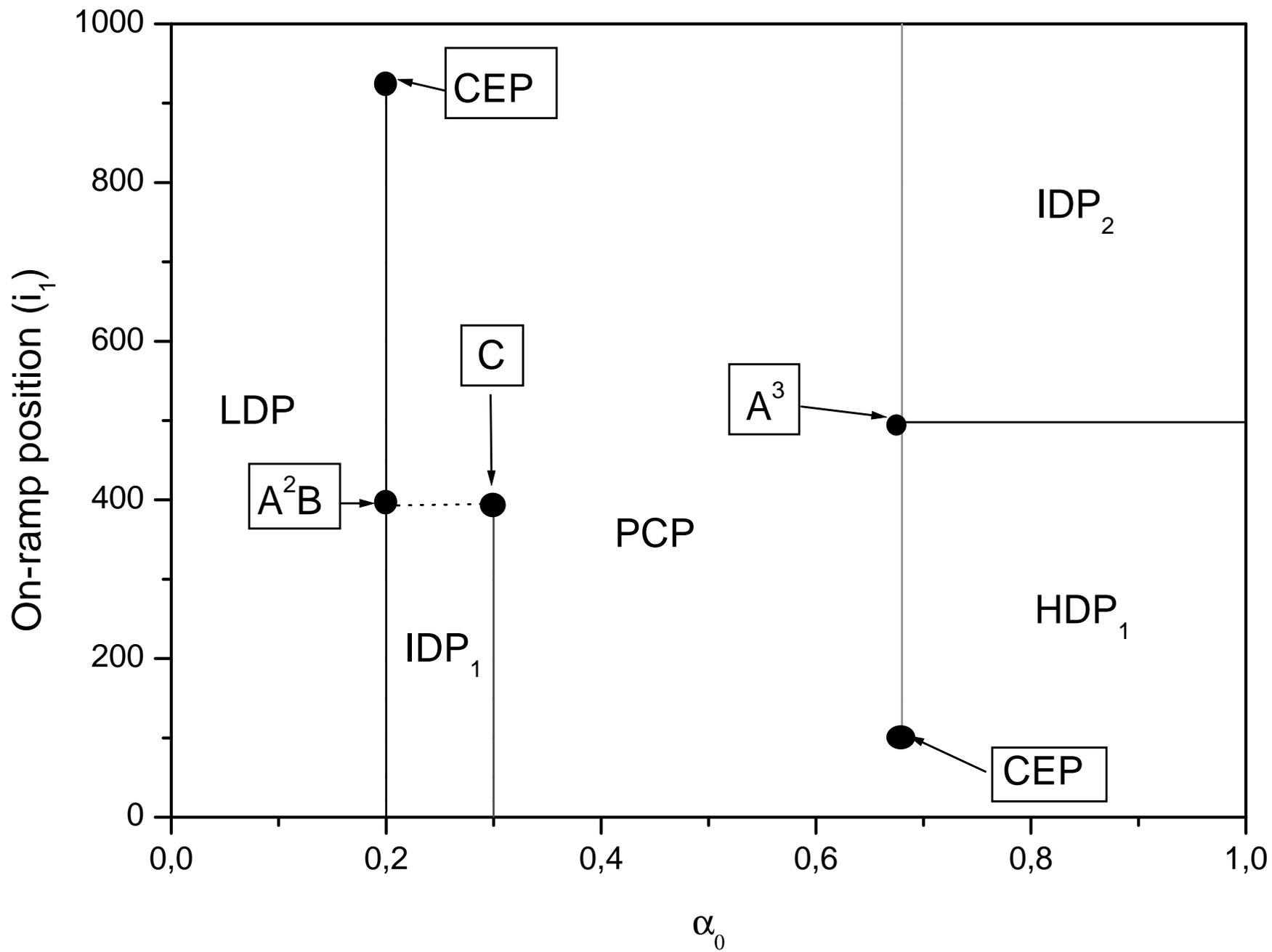
\end{document}